\documentclass[%
 reprint,
nofootinbib,
 amsmath,amssymb,
 aps,
 prd,
]{revtex4-2}
\usepackage{amsmath,amsfonts,amssymb,amsthm,amstext,amscd,eucal,xcolor}
\usepackage{color}
\usepackage[all]{xy}
\usepackage{hyperref}
\usepackage{epsfig}
\usepackage[active]{srcltx}

\usepackage{graphicx}
\usepackage{dcolumn}
\usepackage{bm}

\newcommand{\bea}{\begin{eqnarray}}
\newcommand{\eea}{\end{eqnarray}}

\begin{document}

\title{Killing tensors of generalized Lense-Thirring spacetime}

\author{Saeedeh Sadeghian}
 \email{s.sadeghian@umz.ac.ir}

\affiliation{Department of Theoretical Physics, Faculty of Science, University of Mazandaran, P.O. Box 47416-95447, Babolsar, Iran}%
\affiliation{ICRANet-Mazandaran, University of Mazandaran, P. O. Box 47415-416, Babolsar, Iran.}%

\begin{abstract}
\noindent
We investigate the Hamilton–Jacobi equation of a probe particle moving on $d$-dimensional generalized Lense-Thirring metric.
This spacetime is different from the slowly rotating Myers-Perry black hole at second order in rotation parameters.
We show that the dynamics of the probe particle along the timelike geodesic of the generalized Lense-Thirring spacetime is superintegrable and has more constants of motion with respect to the same dynamics on Myers-Perry black hole. We also discuss the second rank Killing tensors associated with the generalized Lense-Thirring metric. 
\end{abstract}
\setcounter{footnote}{0}

\maketitle
\section{Introduction}
The spacetime outside a slowly rotating massive object is described by the Lense-Thirring metric \cite{Lense-Thirring}. This metric is also given by the slow rotation limit of a Kerr black hole to first order in rotation parameter. Hence, this metric is an approximate solution to the field equation of the pure Einstein theory.  
Another description for this spacetime is that it appears as the asymptotic (large $r$) behavior of the Kerr metric.
Since stationary, this spacetime does not follow the four-dimensional uniqueness theorem. Therefore, different rotating objects can be described by Lense-Thirring metric asymptotically. So, we can appropriately add a coordinate-dependent term of second order in the rotation parameter to the metric which makes the analysis of the new metric simpler\cite{Baines:2020unr,Baines:2021qaw}. Although this change gives a different spacetime, it still satisfies Einstein's equation to the first order in rotation parameter.

This spacetime inherits the rich symmetry structure from the Kerr black hole. The study of geodesic equations on Lense-Thirring spacetime shows that the dynamics of the probe particle along the timelike geodesic is integrable\cite{Baines:2021qaw} and one constant of motion is associated with the nontrivial Killing tensor of the second rank. This is similar to the Carter constant \cite{Carter:1968ks} for the probe particle dynamics on a Kerr black hole.

The Killing tensors of rank-$n$ which satisfy the following equation,
\bea\label{KT-eq}
\nabla^{\{\mu} K^{\nu_1 \cdots \nu_n\}}=0\,,
\eea
 are the generators of some symmetry actions on the phase space of the probe particle of the curved spacetime, as the generalization of the Killing vectors. However, the projection of their action on the configuration space is null, unlike the action of Killing vectors. In this sense, they are associated with \emph{hidden} symmetries.
There are some cases in which the second rank Killing tensor of the $d$-dimensional spacetime, $K^{ab}$, is written as the ``square'' of a Killing-Yano tensor, $Y$, of rank $(d-2)$, i.e. 
\bea
K=Y\cdot Y\,; \quad \nabla_{(\mu} Y_{\nu)\nu_1 \cdots \nu_{d-3}}=0\,.
\eea
Here ($\cdot$) means the contraction of the $(d-3)$ number of the indices. In such cases, the integrability of the Klein-Gordon equation is guaranteed by the integrability of the geodesic equation. 
Then, the Killing-Yano tensor, $Y$, can be written as the Hodge dual of a second rank closed conformal Killing tensor $h$,
\bea
Y=\star h\,,
\eea
 the so-called principal tensor\cite{Frolov:2006dqt}. This can be locally given by a 1-form potential $b$ through 
 \bea \label{potential-b-def}
 h=db\,. 
 \eea 
The existence of principal tensor $h$, plays a crucial role in the integrability of other probes like Maxwell field on the curved spacetime which entails hidden symmetries \cite{Krtous:2006qy,Kubiznak:2007kh,Cariglia:2011qb,Frolov:2006pe,Krtous:2018bvk,Lunin:2017drx} (see \cite{Frolov:2017kze} for recent review). 

The generalization of the four-dimensional Lense-Thirring metric has been studied by adding Maxwell field, cosmological constant, or by going to the higher dimensions\cite{Gray:2021toe}. This study shows that in higher dimensions some new hidden symmetries emerge as a result of the slow rotation limit. 

In the present work, we start with a brief review of the derivation of $d$-dimensional generalized Lense-Thirring spacetime from the slowly rotating Myers-Perry black hole. Then, we develop the hidden symmetry study in \cite{Gray:2021toe} by investigating the Hamilton–Jacobi equation of the probe particle. We show that this equation is separable in an appropriate coordinate system. Using the separation constants, we read the associated Killing tensors of the generalized Lense-Thirring spacetime in $d=2m+1+\epsilon$ dimensions. The remarkable property of this dynamics is that a U$(m)$ symmetry emerges on the $t,r=$constant subspace. Therefore, in addition to the mentioned separation constant, there are more independent constants associated with the second rank Casimirs of this U$(m)$ symmetry which renders the dynamics superintegrable. We present the explicit form of the associated second rank Killing tensors and discuss how their number is restricted by the phase space dimension of the probe particle moving along the timelike geodesic.

\section{Generalized Lense-Thirring metric}
\subsection{ In four dimensions}
Expanding the Kerr metric in rotation parameter ($a$), while keeping mass parameter finite, we have an approximate solution of the Einstein theory up to order $O(a^2)$:
\bea\label{Lenze-O-1}
ds^2&=&-f(r) dt^2+\frac{dr^2}{f(r)}+2a\sin^2\!\theta(f(r)-1)dt  d\phi\nonumber\\
&&+r^2\sin^2\!\theta d\phi^2 +r^2 d\theta^2+O(a^2)\,,
\eea
where the metric function $f(r)$ is given by
\bea\label{f}
f(r)=1-\frac{2M}{r}\,. 
\eea

The Killing tensor of the approximate solution is given by 
\bea
K=2\ a\ \partial_t \ \partial_\phi +\frac{1}{\sin^2\!\theta}(\partial_\phi)^2 +(\partial_\theta)^2+O(a^2)\,.\label{K4}
\eea
The first term is trivial in the sense that it is the tensor product of two Killing vectors, $\partial_t$ and $\partial_\phi$. 
The spherical symmetry can be seen in the slow rotation limit ($a \to 0$) as the second and third term of the above Killing tensor is the metric on a 2-sphere.
Based on \eqref{potential-b-def}, the associated principal tensor is generated by
\bea
2\ b={r^2}dt-{ar^2\sin^2\!\theta}d\phi +O(a^2)\,.\label{b4}
\eea

Adding a term of order $a^2$ to the metric to complete the square of the $t-\phi$ part of the line element, we get the following metric:
\bea\label{Lens-Th-4d}
&&ds^2=-f(r)\, dt^2+\frac{dr^2}{f(r)}+r^2\sin^2\!\theta\Bigl(d\phi \!+\!\frac{a(f(r)-1)}{r^2}dt\Bigr)^2\!\!\nonumber\\
&&\hspace{12mm}+\ r^2 d\theta^2\,,
\eea
where the function $f(r)$ is given by \eqref{f}.
This is also a solution to the Einstein's equations to the linear order in $a$. The advantage of adding the $O(a^2)$ term to the metric is that its analysis is easier than the one in \eqref{Lenze-O-1} \cite{Baines:2020unr,Baines:2021qaw,Gray:2021toe}. Also, the truncation of the Killing tensor in \eqref{K4} satisfies the exact Killing tensor equation.

\subsection{In higher dimensions}
We start from the rotating Myers-Perry black hole \cite{Myers:1986un} in $d=2m+1$ dimensions,
\bea\label{odd-MP}
&&ds^2=-dt^2+\sum_{i=1}^{m}(r^2+a_i^2) \left(d\mu_i^2+\mu_i^2 d\phi_i^2\right)+\frac{\Pi F}{\Pi-2M r^2}dr^2\nonumber\\
&&\hspace{1cm}+\frac{2M r^2}{\Pi F}\left( dt+\sum_{i=1}^{m} a_i \mu_i^2 d\phi_i\right)^2\,,
\eea
where $\mu_i$ are azimuthal coordinates, restricted by $\sum_{i=1}^{m} \mu_i^2=1$. Here, $M$ and $a_i$ are the mass and rotation parameter, respectively and the metric functions $F$ and $\Pi$ are defined through
\bea
\Pi=\prod_{i=1}^{m}(r^2+a_i^2),\hspace{.4cm} F=1-\sum_{i=1}^{m} \frac{a_i^2 \mu_i^2}{r^2+a_i^2}
\,.
\eea
The isometries of this spacetime are generated by $\partial_t$ and $\partial_{\phi_i}$. This metric also admits the principal tensor generated by the following potential \cite{Frolov:2006dqt}:
\bea\label{b-MP}
\hspace{-4mm}2b=\left(r^2+\sum_{i=1}^{m}a_i^2\mu_i^2\right)dt+\sum_{i=1}^ma_i\mu_i^2\left(r^2+a_i^2\right)d\phi_i\,. \
\eea
Taking the slow rotation limit to linear order, we have
\bea
&&ds^2=-\left(1-\frac{2M}{r^{2m-2}}\right)dt^2+r^2 \sum_{i=1}^{m} \left(d\mu_i^2+\mu_i^2 d\phi_i^2\right)\nonumber\\
&&\hspace{1cm}+\frac{dr^2}{1- 2M/r^{2m-2}}+\frac{4M}{r^{2m-2}} \sum_{i=1}^{m} a_i \mu_i^2 d\phi_i dt +O(a_i^2)\,.\nonumber\\
\eea
The same procedure as in the four-dimensional case leads to the generalized Lense-Thirring metric in $d=2m+1+\epsilon$ dimensions\cite{Gray:2021toe},
\bea\label{LTHD}
&&ds^2=-f(r)dt^2+\frac{dr^2}{f(r)}+r^2\sum_{i=1}^m \mu_i^2\Bigl(d\phi_i+\frac{2Ma_i}{r^{2m+\epsilon}} dt\Bigr)^2\nonumber\\
&&\hspace{1cm}+r^2\,\sum_{i=1}^{m+\epsilon}\!d\mu_i^2\!\,,\hspace{-1.2cm}
\eea
where $\epsilon=0,1$ for odd/even spacetime dimensions, respectively. The metric function $f(r)$ is given by
\bea
f(r)\!=\!1-\frac{2M}{r^{2m-2+\epsilon}}
\,.\ 
\eea
The restriction on $\mu_i$ coordinates is
\bea\label{mu}
\sum_{i=1}^{m+\epsilon}\mu_i^2=1\,,
\eea
which hints at the spherical symmetry on a submanifold spanned by $\mu_i$. The Killing vectors of this spacetime are $\partial_t$ and $\partial_{\phi_i}$. Later, we discuss the hidden symmetries associated with the Killing tensors of the above metric.

Just like its four-dimensional counterpart, the metric introduced in \eqref{LTHD} is an approximate solution to the Einstein's equations to the linear order in $a_i$.

\section{Hamilton–Jacobi equation of the probe particle}
For the case of $\epsilon=0$, the restriction \eqref{mu} is written as
\bea
\sum_{I=1}^{m-1}\mu_I^2+\mu_m^2=1\,.
\eea
We solve this for $\mu_m$ and rewrite the metric \ref{LTHD} in terms of $\mu_I$ with $I \in \{1,\cdots,m-1\}$. Then, the Hamilton–Jacobi equation for the probe particle,
\bea\label{geodesic-eq}
-m_0^2=g^{ab}p_a p_b\,,
\eea
moving on this background metric is given by
\bea\label{eq1-odd}
&&\hspace{-1.2cm}-m_0^2=-f(r)^{-1}\left(p_t-\sum_{i=1}^{m}\frac{2 M a_i}{r^{2m}}\,p_{\phi_i}\right)+f(r)\,p_r^2\nonumber\\
&&+r^{-2}\left(\sum_{I,J=1}^{m-1}h^{IJ}\,p_{\mu_I}p_{\mu_J}+\sum_{i=1}^{m}\mu_i^{-2}\,p_{\phi_i}^2\right)\,,
\eea
where 
\bea
h^{IJ}=\delta_{IJ}-\mu_I \mu_J\,.
\eea
Since the metric is stationary and axisymmetric along $\phi_i$, the energy and angular momentum of the point particle on this background metric are conserved and $p_t$, $p_{\phi_i}$'s are some constants. If we define
\bea
\mathcal{U}(r)\equiv f(r)^{-1}\left(p_t-\sum_{i=1}^{m}\frac{2 M a_i}{r^{2m}}\,p_{\phi_i}\right)-f(r)\,p_r^2\,,
\eea
then the $(t,r,\phi_i)$ part of Eq. \eqref{eq1-odd} separates from the rest,
\bea\label{radial-part}
r^2\left(\mathcal{U}(r)-m_0^2\right)=\mathcal{C}\,,
\eea
by introducing a separation constant $\mathcal{C}$.
The $\mu_i$-dependent part of Eq. \eqref{eq1-odd} is 
\bea\label{non-sep-odd}
&&\left(1-\sum_{K=1}^{m-1}\mu_K^{2}\right)\,\sum_{I,J=1}^{m-1}\left(p_{\mu_I}^2-\mu_{I}\,\mu_J\,p_{\mu_I}\,p_{\mu_J}+\mu_I^{-2}\,p_{\phi_I}^2\right)\nonumber\\
&&\hspace{1cm}+\sum_{I=1}^{m-1}\mathcal{C}\,\mu_I^2=\mathcal{C}-p_{\phi_m}^2\,.
\eea
Obviously, this equation is not separable in $\mu_I$ coordinates. Then, we change to $\theta_i$ coordinates in which the equation \eqref{non-sep-odd} is separable,
\bea\label{mu-to-theta}
\mu_m=\cos{(\theta_m)}\,,\quad \mu_I=\hat{x}_I\sin{(\theta_m)}\,,
\eea
where $\hat{x}_I$ is constrained by
\bea
\sum_{I=1}^{m-1}\hat{x}_I^2=1\,.
\eea
To solve this, we introduce another $\theta$ coordinate, $\theta_{m-1}$ and so on. Therefore, we need $m-2$ number of $\theta_i$'s (with $i\in \{3,m\}$) which ranges in $[0,\pi]$ plus one azimuthal angle $\theta_2=\tilde{\phi} \in [0,2\pi]$.
In this coordinate, the $\theta_i$ part of Eq. \eqref{geodesic-eq} becomes separable,
\bea \label{K-const-odd}
p_{\theta_i}^2+\frac{p_{\phi_i}^2}{\cos{\theta_i}^2}+\frac{\mathcal{K}_{(i-1)}}{\sin{\theta_i}^2}=\mathcal{K}_{(i)}\,,\quad i\in\{2,m\}\,,
\eea
where $\mathcal{K}_{(i)}$ are some separation constants and $\mathcal{K}_{(1)}=p_{\phi_1}^2$. 
We note that $\mathcal{K}_{(m)}$ is nothing but the separation constant $\mathcal{C}$, introduced in \eqref{radial-part} for the radial part. 
Up to here, the existence of $(m-1)$ number of $\mathcal{K}_{(i)}$'s, in addition to $m_0, p_t , p_i$, makes the dynamics along the geodesic integrable. However, there are more constants of motion associated with the the generators of U$(m)$ symmetry which will be discussed in the next section (these are similar to the hidden constants of motion introduced in \cite{Hakobyan:2011ir,Galajinsky:2013mla}). 
\section{Hidden symmetries and Killing tensors}
Using the recursion relation between the constants $\mathcal{K}_{(i)}$'s in Eq. \eqref{K-const-odd}, we read the related Killing tensors,
\bea\label{K-k}
&&K_{(k)}^{a b}\partial_a\partial_b=\partial_{\theta_{k}}^2+\frac{\partial_{\phi_{k}}^2}{(\cos{\theta_k})^2}+\left(\prod_{l=0}^{k-2}\sin{\theta_{k-l}}\right)^{-2}\partial_{\phi_{1}}^2+\nonumber\\
&&\hspace{1cm}+\sum_{q=0}^{k-3}\left[\prod_{l=0}^{q}\sin{\theta_{k-l}}\right]^{-2} \left[\partial_{\phi_{k-q-1}}^2+\frac{\partial_{\phi_{k-q-1}}^2}{\left(\cos{\theta_{k-q-1}}\right)^{2}}\right]\,,\nonumber\\
\eea
for $3\le k \le m$ and
\bea
K_{(2)}=\partial_{\theta_2}^2+\cos{\theta_{2}}^{-2}\partial_{\phi_2}^2+\sin{\theta_{2}}^{-2}\partial_{\phi_1}^2\,.
\eea
Furthermore, the generalized Lense-Thirring metric \eqref{LTHD} admits some additional Killing tensors associated with the mentioned U$(m)$ symmetry. To write them explicitly, we first introduce coordinates $x^i,y^i$ in which the mentioned U$(m)$ symmetry is more clear as in \cite{Galajinsky:2013mla},
\bea
x^i=\mu_i\,\cos{(\phi_i)}\,,\qquad y^i=\mu_i\,\sin{(\phi_i)}\,.
\eea
Here, $i$ runs from 1 to $(m-1)$. If one writes the reduced metric on $(\mu_i-\phi_i)$ subspace in complex coordinates $z^j=x^j+i\, y^j$, 
\bea
ds^2\Big|_{t,r=const.}=d\mu_i^2+\mu_i^2 d\phi_i^2=dz_id\bar{z}_i\,,
\eea
then the U$(m)$ symmetry manifests.

The second rank Casimir of the U$(m)$ symmetry, $I_{ij}$, is given by
\bea
I_{ij}=-\frac14(\xi_{ij}^2+\rho_{ij}^2)\,, \qquad (i<j)
\eea
in which the vector fields $\xi_{ij}$ and $\rho_{ij}$ are defined by
\bea
&&\xi_{ij}\equiv x^{i}\frac{\partial}{\partial{x^{j}}}- x^{j}\frac{\partial}{\partial{x^{i}}}+ y^{i}\frac{\partial}{\partial{y^{j}}}- y^{j}\frac{\partial}{\partial{y^{i}}}\,, \nonumber\\
&&\rho_{ij}\equiv x^{i}\frac{\partial}{\partial{y^{j}}}- y^{j}\frac{\partial}{\partial{x^{i}}}+ x^{j}\frac{\partial}{\partial{y^{i}}}- y^{i}\frac{\partial}{\partial{x^{j}}}\,.
\eea
The vector $\rho_{ij}$ is symmetric under the exchange of $i,j$ and $\xi_{ij}$ is antisymmetric under this exchange. As $\xi_{ij}$ does not contribute to the diagonal components of $I_{ij}$,  it simplifies considerably. A simple algebra shows that $I_{ii}$ is the trivial Killing tensor since $I_{ii}=\partial_{\phi_i} \partial_{\phi_i}$ (there is no summation on the repeated indices). Therefore, the nontrivial Killing tensors in $\mu_i$ basis are given by 
\bea
\hspace{-.4cm}-4\ I_{ij}=(\mu_i\partial_{\mu_j}-\mu_j\partial_{\mu_i})^2+\left(\frac{\mu_j}{\mu_i}\partial_{\phi_i}+\frac{\mu_i}{\mu_j}\partial_{\phi_j}\right)^2,\ 
\eea
for $i<j$. 
However, \emph{all} of the $I_{ij}$ components (when $i<j$) do not lead to functionally independent constants of motion for the  Hamilton–Jacobi equation. In \cite{Galajinsky:2013mla}, it has been shown that the constants associated with $I_{ij}$ not only includes $(m-1)$ number of $\mathcal{K}_{(i)}$ but also it contains $(m-2)$ number of new \emph{independent} constants of motion constructed out of $I_{(i-1)i}$. The explicit form of $I_{(i-1)i}$ in $\theta_i$ coordinates is given by
\bea\label{extra-KT}
&&I_{(i-1)i}=\left(\sin{\theta_{i-1}}\cot{\theta_i}\frac{\partial}{\partial{\theta_{i-1}}}-\cos{\theta_{i-1}}\frac{\partial}{\partial{\theta_{i}}}\right)^2\nonumber\\
&&\hspace{1.2cm}+\left(\cos{\theta_{i-1}\tan{\theta_i}\frac{\partial}{\partial{\phi_{i}}}+\frac{\cot{\theta_i}}{\cos{\theta_{i-1}}}}\frac{\partial}{\partial{\phi_{i-1}}}\right)^2\,, \ \ \ \
\eea
where $i \in\{2,m\}$. One can explicitly check that it satisfies the Killing tensor equation \eqref{KT-eq}.

\section{Discussion}
In this work, we studied the Hamilton–Jacobi equation of the probe particle on the generalized Lense-Thirring metric in $d=2m+1$ dimensions. This metric is the solution to the pure Einstein theory to the linear order in the rotation parameter $a_i$. It would be interesting to analyze the geodesic of the extended Lense-Thirring as the solution to the other theories such as Einstein-Maxwell-Lambda to the first order in rotation parameter\cite{Gray:2021toe, Kubiznak:2022vft, Gray:2021roq}.

We observed that the dynamics of the probe particle along the timelike geodesic of the generalized Lense-Thirring spacetime in $d$-dimensions is superintegrable. 
For a system with $(m-1)$ degrees of freedom, it is maximally superintegrable if it has a $2(m-1)-1$ number of independent constants of motion. This is the case for the reduced phase space related to the independent $\mu_i$'s as $(m-1)$ number of $\mathcal{K}_{(i)}$'s and $(m-2)$ number of constants associated with $I_{i(i-1)}$'s renders the dynamics maximally superintegrable.

Regarding the principal tensor ($h$), one idea is to start from the principal tensor associated with Myers-Perry black hole\eqref{b-MP} and take the slow rotation limit which gives 
\bea\label{h-Kerr}
h=d\,b\,,\qquad b=\frac{r^2}{2}\left(dt+\sum_{i=1}^{m} a_i\,\mu_i^2 d\phi_i\right)\,,
\eea
to linear order in $a_i$. 
A straightforward calculation shows that the Killing tensor associated with \eqref{h-Kerr} \emph{in the slow rotation limit} is 
\bea\label{KT-standard}
&&K^{ab}\partial_a\partial_b=2\sum_{i=1}^m a_i \partial_t \partial_{\phi_i}\nonumber\\
&&\hspace{1.2cm}+\sum_{I,J=1}^{m-1}(\delta_{IJ}-\mu_I\mu_J)\partial_{\mu_I} \partial_{\mu_J}+\sum_{i=1}^m \frac{\left(\partial_{\phi_i}\right)^2}{\mu_i^{2}}\,,\ \ \
\eea
in odd dimensions. Here, we imposed the restriction \eqref{mu}, so that the indices of $\partial_\mu$ runs from 1 to $(m-1)$.
Therefore, the second term is the metric on an $m$-sphere. Changing the bound of summation in \eqref{h-Kerr} does not yield a different result for the Killing tensor other than \eqref{KT-standard} \emph{to the linear order in $a_i$}. An interesting question is, are the Killing tensors $I_{ij}$ constructed of some principal tensors?
We will come back to this question in the future.

Finally, we note that the discussion in even dimensions is very similar to the odd-dimensional case. To avoid repetition, we left the details in the Appendix. 
%

\section*{Acknowledgments}
I am grateful to M.M. Sheikh-Jabbari to bringing to my attention the results of \cite{Gray:2021toe} which lead to the current project. The author also acknowledges the collaboration of H. Golchin and H. Demirchian in the early stages of this research. I would like to thank D. Kubiznak for his comments on the draft. The support of ICTP program network scheme NT-04 is appreciated.

\appendix
\section{Geodesic equation in even dimensions}\label{even-d}
In even dimensions ($\epsilon=1$) , the restriction \eqref{mu} reads
\bea\label{mu-nu}
\sum_{i=1}^{m}\mu_i^2+\nu^2=1\,,
\eea
where, for convenience, we replaced $\mu_{m+1}$ by $\nu$. This can be solved for $\nu$, then the metric reduces to
\bea
&&ds^2=-f(r)dt^2+\frac{dr^2}{f(r)}+r^2\sum_{i=1}^m \mu_i^2\Bigl(d\phi_i+\frac{2Ma_i}{r^{2m+1}} dt\Bigr)^2\nonumber\\
&&\hspace{1.2cm}+\ r^2\sum_{i,j=1}^m h_{ij}\,\!d\mu^i\,d\mu^j\,,
\eea
where the metric functions are
\bea
\!\!\!\ f(r)\!=\!1-\frac{2M}{r^{2m-1}}+\frac{r^2}{\ell^2}\,, \quad h_{ij}=\delta_{ij}+\frac{\mu_i\,\mu_j}{\nu^2}\,,
\eea

The Hamilton–Jacobi equation of the probe particle on this background metric is  given by
\bea\label{eq1}
&&\hspace{-1.2cm}-m_0^2=-f(r)^{-1}\left(p_t-\sum_{i=1}^{m}\frac{2 M a_i}{r^{2m+1}}\,p_{\phi_i}\right)+f(r)\,p_r^2\nonumber\\
&&+r^{-2}\left(\sum_{i,j=1}^{m}h^{ij}\,p_{\mu_i}p_{\mu_j}+\sum_{i=1}^{m}\mu_i^{-2}\,p_{\phi_i}^2\right)\,,
\eea
where $h^{ij}=\delta_{ij}-\mu_i \mu_j\,$. 
Since the metric is stationary and axisymmetric along $\phi_i$'s the energy and angular momentum of the point particle on this background metric are conserved and $p_t$, $p_{\phi_i}$'s are some constants. If we define
\bea
\hspace{-.4cm}\mathcal{U}(r)\equiv f(r)^{-1}\left(p_t-\sum_{i=1}^{m}\frac{2 M a_i}{r^{2m+1}}\,p_{\phi_i}\right)-f(r)\,p_r^2\,,\,
\eea
then the $r$-dependent part separates from the rest of Eq. \eqref{eq1} by introducing the separation constant $\mathcal{C}$,
\bea\label{r-part-sep}
r^2\left(\mathcal{U}(r)-m_0^2\right)=\mathcal{C}\,.
\eea
The $\mu_i$ part of Eq. \eqref{eq1},
\bea\label{non-sep-even}
\sum_{i,j=1}^{m}\left(p_{\mu_i}^2+\mu_{i}\,\mu_j\,p_{\mu_i}\,p_{\mu_j}+\mu_i^{-2}\,p_{\phi_i}^2\right)=\mathcal{C}\,,
\eea
 is not separable in $\mu_i$ coordinates.   
 
Then, we change the coordinates to $\theta_i$ coordinates,
\bea
\hspace{-.7cm}\nu=\cos{(\theta_{m+1})}\,,\ \ \mu_i=\hat{x}_i\sin{(\theta_{m+1})}\,;\ \ \sum_{i=1}^{m}\hat{x}_i^2=1\,.\ \ \ \ 
\eea
In these coordinates, the $\theta_i$ part of the Hamilton–Jacobi equation \eqref{geodesic-eq} is written as
\bea
\hspace{-.6cm}p_{\theta_i}^2+\frac{p_{\phi_i}^2}{\cos{\theta_i}^2}+\frac{\mathcal{K}_{(i-1)}}{\sin{\theta_i}^2}=\mathcal{K}_{(i)}\,,\quad i\in\{2,m+1\}\,, \ \ \ \ \
\eea
with
\bea
p_{\phi_{m+1}}=0\,,\quad \mathcal{K}_{(1)}=p_{\phi_1}^2\,.
\eea
The Killing tensors related to these constants are
\bea
&&K_{(k)}^{a b}\partial_a\partial_b=\partial_{\theta_{k}}^2+\frac{\partial_{\phi_{k}}^2}{(\cos{\theta_k})^2}+\sum_{q=0}^{k-3}\left(\prod_{l=0}^{q}\sin{\theta_{k-l}}\right)^{-2}\!\!\!\!\partial_{\theta_{k-q-1}}^2\nonumber\\
&&\hspace{1.5cm}+\sum_{q=0}^{k-3}\frac{\left(\prod_{l=0}^{q}\sin{\theta_{k-l}}\right)^{-2}}{\left(\cos{\theta_{k-q-1}}\right)^{2}}\partial_{\phi_{k-q-1}}^2\nonumber\\
&&\hspace{1.5cm}+\left(\prod_{l=0}^{k-2}\sin{\theta_{k-l}}\right)^{-2}\partial_{\phi_{1}}^2\,,
\eea
where $(2\le k \le m+1)$ while $\mathcal{K}_{(m+1)}$ is equal to $\mathcal{C}$ which appeared as the separation constant for the radial part, in Eq. \eqref{r-part-sep}.


\end{document}